\documentclass[12pt]{article}
\usepackage{epsfig, amssymb}
\usepackage{graphicx,epsfig}
\usepackage{color} 
\begin{document}
\begin{titlepage}
\thispagestyle{empty}
\begin{flushright}
\end{flushright}

\bigskip

\begin{center}
\noindent{\Large \textbf
{Electric-magnetic duality implies (global) conformal invariance}}\\ 
\vspace{2cm} \noindent{Sung-Pil Moon\footnote{mlnow@naver.com}, Sang-Jin Lee\footnote{lee3jjang@naver.com}, Ji-Hye Lee\footnote{elaia414@hanyang.ac.kr} and
Jae-Hyuk Oh\footnote{e-mail:jack.jaehyuk.oh@gmail.com}}

\vspace{1cm}
  {\it
Department of Physics, Hanyang University, Seoul 133-791, Korea\\
 }
\end{center}

\vspace{0.3cm}
\begin{abstract}
We have examined quantum theories of electric magnetic duality invariant vector fields enjoying classical conformal invariance in 4-dimensional flat spacetime. We extend Dirac's argument about ``the conditions for a quantum field theory to be relativistic" to ``those for a quantum theory to be conformal". We realize that electric magnetic duality invariant vector theories together with classical conformal invariance defined in 4-$d$ flat spacetime are still conformally invariant theories when they are quantized
in a way that electric magnetic duality is manifest. 
\end{abstract}
\end{titlepage}


\section{Introduction}
Electric magnetic duality is originally observed from Maxwell equations, which describe one of the fundamental forces in nature. Under switching $\vec E\rightarrow \vec B$ and $\vec B \rightarrow -\vec E$, where $\vec E$ is electric field and $\vec B$ is magnetic field (without considering any electric and magnetic sources), the Maxwell equations are invariant\cite{Deser1}. The duality is extended to string theory and various kinds of field theories of free massless fields with various spins, sometimes to those in curved spacetime e.g. Maxwell system in de Sitter spacetime and to approximate non-Abelian dualities \cite{Henneaux:2004jw,Deser:2004xt,Deser:2013xb,Deser:2005sz,Jatkar:2012mm}.

One of the interesting directions of developing electric magnetic duality is a research if electric magnetic duality ensures that certain classical symmetry of a system is retained when the system is quantized(e.g. see \cite{Bunster:2012hm}). In \cite{Bunster:2012hm}, the authors argue that electric magnetic duality can ensure if a classical vector field theory enjoying Lorentz symmetry is still Lorentz invariant even when it is quantized.

The pioneering argument started from a paper by Dirac\cite{Dirac1} in 1962. In his paper, he discussed this issue as follows. It is not manifest if a quantum field theory keeps its classical symmetry(symmetry of the classical Lagrangian and equations of motion) because of (e.g.) the ordering issue of the field variables(due to the second quantization rule on them). Since a state in quantum field theory can change to another representation by unitary transform and its dynamics is described by unitary time evolution, acting symmetry generators(spatial translation, rotation and boost, temporal translation) on that state, then if the second quantization is consistent with the algebra of the symmetry generators, then this ensures that the symmetry retains in its quantum field theory. 

More precisely, he introduces a canonical pair of quantum fields as $\xi$ and $\eta$ satisfying
\begin{equation}
[\xi,\eta^\prime]=\delta,
\end{equation}
where prime denotes that the field variable depends on prime coordinate i.e. $\eta^\prime=\eta(x^\prime)$, $\delta=\delta^d(x-x^\prime)$, $d$-dimensional $\delta$-function and so it is an equal-time commutator
\footnote{For further discussion, even if we develop every mathematical equation in terms of $d$, in fact we restrict ourselves to $d=3$ case only.}
. $\xi$ may become a field variable in the theory and $\eta$ is its canonical conjugate. From them, he constructs a momentum density $K_s$ and introduces an energy density $U$, which provide the representation of the symmetry generators,
where the index $s$ is space index\footnote{We will use $s,t,r,u$ to be spatial indices running from 1 to 3.}. It turns out that such symmetry generators constructed from $K_s$ and $U$ satisfy Poincare algebra if the energy density satisfies the following commutation relation:
\begin{equation}
\label{intri-uu}
[U,U^\prime]=K_{t,t}\delta+2K_t\delta_{,t},
\end{equation}
where $A_{,s}\equiv \frac{\partial A}{\partial x^s}$.

By using this observation, the authors in \cite{Bunster:2012hm} discovered the following: Suppose a vector field theory in 4-$d$ flat spacetime which enjoys electric magnetic duality and Lorentz symmetry is quantized in a way that electric magnetic duality is manifest, more precisely it is requested for its second quantization rule to be
\begin{equation}
\label{BB-commm}
[\mathcal B^a_s,\mathcal B^b_t]=\epsilon^{ab}\epsilon_{stu}\delta_{,u},
\end{equation}
where $a,b=1,2$ are $SO(2)$ indices related to electric magnetic duality rotation, $\epsilon$ is fully anti-symmetric tensor, $ \vec \mathcal B^1=\vec E$ and $ \vec\mathcal B^2=\vec B$. One can define the momentum density and the energy density from the fields $\mathcal B^a_s$ as
\begin{equation}
\label{em-density-intro}
K_r=-\frac{1}{2}\mathcal B^{as}\mathcal B^{bt}\epsilon^{ab}\epsilon_{str} {\rm \ \ and \ \ } U=f(h,v), 
\end{equation}
where
\begin{equation}
h=\frac{1}{2}\mathcal B^{as}\mathcal B^{bt}\delta^{ab}\delta_{st}, {\ \ }v=K_rK^r
\end{equation}
and $f(h,v)$ satisfies the following condition
\begin{equation}
(f_{,h})^2+4f_{,h}f_{,v}+4(f_{,v})^2=k,
\end{equation}
for some constant $k$. The momentum density generates Lie derivative along a spatial vector field $v_i$ as $\mathcal L_v \Phi(\mathcal B)=[\Phi,\int d^dx v^s K_s]$ for some field $\Phi$.  It turns out that such an energy density satisfies the commutation relation that Dirac suggested in his paper. Therefore, one can find out that the vector field theory is manifestly Lorentz invariant when it is quantized.

In this paper, we have extended such discussion to conformal symmetry. Our motivation is that $U(1)$ vector field theory in $4-d$ flat spacetime, whose Lagrangian density is comprised of its kinetic term only, is conformally invariant, since its stress energy tensor vanishes. Thus, one may ask {\it if quantum version of such kind of classical field theory is still conformally invariant when its second quantization rule manifestly enjoys electric magnetic duality transform}.

In fact, we have shown that the theory is still conformal by examining conformal algebra with the similar manner that Dirac studied. In section \ref{Conditions for a 4}, we develop the conditions that the momentum and the energy densities satisfy for this. It turns out that the energy density still satisfies (\ref{intri-uu}) and therefore the momentum density and the energy density that Dirac suggested also satisfy conformal algebra under one condition that {\bf conformal dimension of the energy density is $d+1$}. The simplest example for such case is $U=h$.

In section \ref{section2}, we conclude that since a specific class of the energy density (\ref{em-density-intro}) whose conformal dimension is $d+1$  obtained in \cite{Bunster:2012hm} satisfy the same commutation relation (\ref{intri-uu}), then conformal symmetry is retained in such quantum theory of $U(1)$ vector field which is manifestly invariant under electric magnetic duality rotation.
 
The final issue to discuss is central charge. There possibly is conformal anomaly which shows up in OPE's of energy momentum tensors, then transformation rule of the field variables and the momentum and energy densities will be affected by the anomaly. However, as long as we restrict ourselves in global conformal symmetry, central charge cannot affect the transformation rules. For example, in 2-$d$ CFT, the central charge contribution to the transformation of stress energy tensor is given by derivative of the transformation parameters.
\section{Conditions for a 4-$d$ quantum field theory to be conformal}
\label{Conditions for a 4}
In this section, we extend Dirac's argument about conditions for a quantum field theory to retain Poincare symmetry to conformal symmetry.
\paragraph{Conformal algebra}
Conformal algebra in $d+1$-dimensional space time is given by
\begin{eqnarray}
\label{conformal algebra}
[D,P_\mu]&=&-P_\mu, {\ \ } [D, \kappa_\mu]=\kappa_\mu, {\ \ }[\kappa_\mu,P_\nu]=-2(g_{\mu\nu}D+L_{\mu\nu}), \\ \nonumber
[\kappa_\rho,L_{\mu\nu}]&=&(g_{\rho\mu}\kappa_\nu-g_{\rho\nu}\kappa_\mu), {\ \ }[P_\rho,L_{\mu\nu}]=g_{\rho\mu}P_\nu-g_{\rho\nu}P_\mu \\ \nonumber
[L_{\mu\nu},L_{\rho\sigma}]&=&g_{\nu\rho}L_{\mu\sigma}+g_{\mu\sigma}L_{\nu\rho}-g_{\mu\rho}
L_{\nu\sigma}-g_{\nu\sigma}L_{\mu\rho}, {\rm\ \ and \ the\ others\ vanish,}
\end{eqnarray}
where $D$ is dilatation, $\kappa_\mu$ is special conformal, $P_\mu$ is translation and $L_{\mu\nu}$ is rotation and boost generators
\footnote{The generators are given by
\begin{equation}
D=x^\mu P_\mu, {\ \ }L_{\mu\nu}=x_\mu P_\nu-x_\nu P_\mu, {\ \ }\kappa_\mu=2x_\mu x^\nu P_\nu-x^\nu x_\nu P_\mu,
\end{equation}
in terms of translation generator, $P_\mu$.
}. $g_{\mu\nu}$ is $d+1$-dimensional flat spacetime metric, whose signature is chosen as $g_{\mu\nu}={\rm diag}(+,-,-,...,-)$.

The symmetry generators are sorted to two different classes. The first class is a set of the generators having the quantum fields transform in spatial directions and the second class is those forcing them transform in temporal direction. The former provides unitary transform of the fields in a given spacelike hypersurface and the later does dynamics of the fields.

\paragraph{Momentum density}
We first examine the generators having the fields transform in spatial directions.
For this, we decompose these generators into spatial and temporal parts as
\begin{eqnarray}
P_\mu\rightarrow P_s, P_0, {\ \ \ \ }L_{\mu\nu}\rightarrow L_{st}, L_{0t}, \\ \nonumber
\kappa_\mu\rightarrow \kappa_s,\kappa_0{\rm\ \  and\ \ }D\rightarrow D^{(s)}+D^{(t)},
\end{eqnarray}
where we have defined the spatial parts of the symmetry generators in terms of a momentum density, $K_s$ as
\begin{eqnarray} 
P_t&=&\int K_t d^dx, {\ \ \ } L_{rs}= \int (x_r K_s-x_sK_r)d^dx\\ \nonumber
D^{(s)}&=&-\int x_sK_s d^dx, {\ \ \ }\kappa_t=\int({-2x_t x_r K_r+ x_r x_r K_t })d^dx
\end{eqnarray}


To specify field variables, $V_t^{(1)}$, $V_t^{(2)}$ and the momentum density in our vector theory, we introduce variables $\xi_s$ and $\eta_s$ as
\begin{equation}
V^{(1)}_{t}=\eta_t, {\ }V^{(2)}_{t}=\xi_t, {\rm \ and \ }K_t=\eta_u\xi_{u,t}-(\eta_u\xi_t)_{,u},
\end{equation}
where $\eta_s$ and $\xi_s$ form a canonical pair as
\begin{equation}
[\xi_t,\eta_s^\prime]=\delta_{ts}\delta,
\end{equation}
where $\delta_{ts}$ is Kronecker's delta whereas $\delta$ is $d$-dimensional delta function.

By using canonical commutation relation of $\xi_s$ and $\eta_s$, transformation rules of the field variables are obtained as
\begin{eqnarray}
[V^{(1)}_t,P_r]&=&V^{(1)}_{t,r} \\ \nonumber
[V^{(1)}_t,L_{rs}]&=&x_rV^{(1)}_{t,s}-x_sV^{(1)}_{t,r}+(-\delta_{rt}V^{(1)}_s+\delta_{st}V^{(1)}_r) \\ \nonumber
[V^{(1)}_t,D^{(s)}]&=&-x_s V^{(1)}_{t,s}+(\Delta_{1}V^{(1)}_t), \\ \nonumber
[V^{(1)}_t,\kappa_s]&=&-2x_sx_rV^{(1)}_{t,r}+x_rx_rV^{(1)}_{t,s}+(2\delta_{ts}x_rV^{(1)}_r
-2x_tV^{(1)}_s+2\Delta_1x_sV^{(1)}_t)
\end{eqnarray}
and
\begin{eqnarray}
[V^{(2)}_t,P_r]&=&V^{(1)}_{t,r} \\ \nonumber
[V^{(2)}_t,L_{rs}]&=&x_rV^{(1)}_{t,s}-x_sV^{(1)}_{t,r}+(-\delta_{rt}V^{(1)}_s+\delta_{st}V^{(1)}_r) \\ \nonumber
[V^{(2)}_t,D^{(s)}]&=&-x_s V^{(1)}_{t,s}+(\Delta_{2}V^{(1)}_t), \\ \nonumber
[V^{(2)}_t,\kappa_s]&=&-2x_sx_rV^{(2)}_{t,r}+x_rx_rV^{(2)}_{t,s}+(2\delta_{ts}x_rV^{(2)}_r
-2x_tV^{(2)}_s+  2\Delta_2 x_sV^{(2)}_t)
\end{eqnarray}
where $\Delta_1=d-1$ and $\Delta_2=1$, which are conformal dimensions of the field variables, $V^{(1)}_t$ and $V^{(2)}_t$ respectively.

From these we can obtain the following relations:
\begin{eqnarray}
[K_t,P_r]&=&K_{t,r}, \\ \nonumber
[K_t,L_{rs}]&=&x_rK_{t,s}-x_sK_{t,r}-\delta_{rt}K_s+\delta_{st}K_r, \\ \nonumber
[K_t, D^{(s)}]&=&-x_rK_{t,r}+(\Delta_1+\Delta_2+1)K_t \\ \nonumber
[K_t,\kappa_s]&=&-2x_s x_r K_{t,r}+x_r x_r K_{t,s}+2\delta_{st}x_rK_r
+2(\Delta_1+\Delta_2+1)x_s K_t-2x_tK_s,
\end{eqnarray}
which provide the commutation relations of conformal algebra for the spacetime indices $\mu$ and $\nu$ to be restricted in $\mu,\nu=1,2...,d$.

\paragraph{Energy density}
To complete the conformal algebra(\ref{conformal algebra}), we need to examine the temporal parts of the generators. To do this, we define a local quantity, ``energy density" $U$ and express these generators by it as
\begin{eqnarray}
\label{generator-definitionsU}
P_0&=&\int U d^dx,{\ \ }L_{t0}=\int x_t U d^d x,\\ \nonumber
D^{(t)}&=&0,{\ \ }\kappa^{(t)}_0=\int x_sx_sU d^d x.
\end{eqnarray}
This energy density is scalar under spatial parts of symmetry transforms and we suppose that it has conformal dimension $\Delta_E$, so it
might transform as below:
\begin{eqnarray}
[U,P_t]&=&U_{,t},{\ \ }[U,L_{st}]=x_s U_{,t}-x_t U_{,s}, \\ \nonumber
[U,D^{(s)}]&=&-x_s U_{,s}{{+\Delta_E U}}, {\ \ }[U,\kappa_s]=-2x_s x_r U_{,r}+x_r x_r U_{,s}{{+2\Delta_E x_sU}}
\end{eqnarray}
Such energy density commutation relations lead 
\begin{eqnarray}
[P_0,P_t]&=&0,{\ \ }[P_0,L_{st}]=0,{\ \ }[P_s,L_{t0}]=-\delta_{st}P_0,{\ \ }[L_{t0},L_{rs}]=
\delta_{ts}L_{r0}-\delta_{tr}L_{s0} \\ \nonumber
[D^{(s)},P_0]&=&{{-}}P_0,{\ \ }[D^{(s)},L_{0t}]=0,{\ \ }[\kappa_r,L_{0t}]=\delta_{rt}\kappa_0, {\ \ }[\kappa_0,L_{st}]=0\\ \nonumber
[D^{(s)},\kappa_0]&=&\kappa^{(t)}_0,{\ \ }[\kappa_0,P_t]=-2L_{0t}
,{\ \ }[\kappa_s,P_0]=-2L_{s0}, {\ \ }[\kappa_0,\kappa_s]=0,
\end{eqnarray}
under one condition that {\bf the conformal dimension of the energy density}
\begin{equation}
\Delta_E=d+1.
\end{equation}
They are the commutation relations between temporal and spatial parts of the generators. 
Finally, we request the commutation relations between temporal parts of the generators to complete our discussion. They are given by
\begin{eqnarray}
[P_0,P_0]&=&0,{\ \ }[L_{t0},L_{s0}]=L_{st},{\ \ }[P_0,L_{0t}]=P_t,{\ \ }  [\kappa_0,L_{0t}]=\kappa_t,{\ \ }\\ \nonumber
[\kappa_0,P_0]&=&-2D^{(s)},{\ \ }[\kappa_0,\kappa_0]=0,
\end{eqnarray}
These are translated to the following equations by using (\ref{generator-definitionsU}):
\begin{eqnarray}
\label{energy density relation0}
\int\int [U,U^\prime]d^dxd^dx^\prime&=&0 \\
\label{energy density relation1}
\int\int x_tx^\prime_s[U,U^\prime]d^dxd^dx^\prime&=&\int(x_sK_t-x_tK_s)d^dx \\
\label{energy density relation2}
\int\int x_t[U,U^\prime]d^dx d^dx^\prime&=&\int K_t d^dx \\
\label{energy density relation3}
\int\int x_sx_sx^\prime_t[U,U^\prime]d^dx d^dx^\prime&=&\int(2x_tx_sK_s-x_sx_sK_t)d^d x\\
\label{energy density relation4}
\int\int  x_sx_s[U,U^\prime]d^dx d^dx^\prime&=&2\int x_sK_sd^dx \\
\label{energy density relation5}
\int\int x_u x_u x^\prime_s x^\prime_s[U,U^\prime]d^dxd^dx^\prime&=&0
\end{eqnarray} 
The remaining task is to find out commutation relation between energy densities satisfying the above relations. We start with the most general form of the energy density commutation relation as Dirac suggested\cite{Dirac1}. It is
\begin{equation}
\label{commutator-UUprime}
[U,U^\prime]=a\delta+b_r\delta_{,r}+c_{rs}\delta_{,rs}+d_{rst}\delta_{,rst}+...,
\end{equation}
where 
the coefficients in front of $\delta$-functions are functions of $x_s$ only.
If we switch $U$ and $U^\prime$, by its anti commuting nature we have
\begin{eqnarray}
\label{commutator-UprimeU}
[U^\prime,U]&=&a\delta-b^\prime_r\delta_{,r}+c^\prime_{rs}\delta_{,rs}-d^\prime_{rst}\delta_{,rst}+..., \\ \nonumber
&=&a\delta-(b_r\delta)_{,r}+(c_{rs}\delta)_{,rs}-(d_{rst}\delta)_{,rst}+... \\ \nonumber
&=&\delta(a-b_{r,r}+c_{rs,rs}-d_{rst,rst}+...) \\ \nonumber
&+&\delta_{,r}(-b_r+2c_{ru,u}-3d_{rsu,su}+...) \\ \nonumber
&+&\delta_{,rs}(c_{rs}-3d_{rsu,u}+...).
\end{eqnarray}
Since (\ref{commutator-UUprime}) and (\ref{commutator-UprimeU}) are added up to zero, so from that condition we have
\begin{eqnarray}
\label{a-eq}
0&=&2a-b_{r,r}+c_{rs,rs}-d_{rst,rst}+... \\ 
\label{c-eq}
0&=&2c_{rs,s}-3d_{rst,st}+... \\ 
\label{cprime-eq}
0&=&2c_{rs}-3d_{rsu,u}+...\\ \nonumber
...
\end{eqnarray}
(\ref{a-eq}) gives a solution of $a$ as
\begin{equation}
a=\alpha_{r,r}, {\rm \ \ where\ \ }2\alpha_r=b_{r}-c_{rs,s}+d_{rst,st}-...,
\end{equation}
and (\ref{c-eq}) means that $c_{ru,u}$ is indeed second derivative, then
\begin{eqnarray}
\int (2\alpha_r-b_r)d^dx&=&0, {\rm \ \ and\ \ }\int x_s(2\alpha_r-b_r)d^dx=0,\\ \nonumber
{\rm \ \ since\ \ } 2\alpha_r-b_r&=&-c_{rs,s}+d_{rst,st}-...\rightarrow({\rm second\ derivative\ and \ higher} )
\end{eqnarray}
By using these, we derive more useful relations as
\begin{eqnarray}
\label{comm-U-int}
\int [U,U^\prime]d^dx^\prime&=&\alpha_{r,r} \\
\int x_s^\prime[U,U^\prime]d^dx&=&x_s\alpha_{r,r}-b_s
\end{eqnarray}
After all, we plug (\ref{commutator-UUprime}) into (\ref{energy density relation1}-\ref{energy density relation4}) to fix coefficients of $\delta$-functions(and derivative of them) on the
right hand side of (\ref{commutator-UUprime}). The relation(\ref{comm-U-int}) directly solves (\ref{energy density relation0}).
(\ref{energy density relation2}) gives
\begin{eqnarray}
\int K_td^dx=\int x_t \alpha_{r,r}d^dx=\int \alpha_td^dx=\frac{1}{2}\int b_t d^dx,
\end{eqnarray}
where we have used (\ref{comm-U-int}). From this, we get the most general form of the solutions $\alpha_r$ and $\beta_r$ as
\begin{equation}
\label{ansztz--apha-b}
\alpha_t=K_t+\beta_{tr,r}+\zeta_{,t}{\rm\ \ and\ \ }b_t=2K_t+\bar\beta_{tr,r}+\bar\zeta_{,t},
\end{equation}
where $\beta_t$,${\bar\beta_t}$,${\zeta}$ and $\bar\zeta$ are arbitrary functions of $x_s$.
(\ref{energy density relation1}) provides
\begin{eqnarray}
&{\ }&\int(x_sK_t-x_tK_s)d^dx=\int x_t(x_s\alpha_{u,u}-b_s)=\frac{1}{2}\int d^dx(x_sb_t-x_tb_s)\\ \nonumber
&=&\frac{1}{2}\int d^dx(2x_sK_t-2x_tK_s+x_s\bar\beta_{tr,r}-x_t\bar\beta_{sr,r}+x_s\bar\zeta_{,t}-x_t\bar\zeta_{,s}),
\end{eqnarray}
This relation restricts $\bar\beta_{st}$ to be
\begin{equation}
\int(\bar\beta_{ts}-\bar\beta_{st})d^dx=0.
\end{equation}
and similarly
\begin{equation}
\int(\beta_{ts}-\beta_{st})d^dx=0.
\end{equation}
Next, consider (\ref{energy density relation3}), which is given by
\begin{eqnarray}
2\int x_sK_sd^dx=\int x_sx_s\alpha_{t,t}d^dx=\int 2x_t(K_t+\beta_{tr,r}+\zeta_{,t}),
\end{eqnarray}
which provides conditions for $\beta_{st}$ and $\zeta$ as
\begin{equation}
\int(\beta_{tt}+d\zeta) d^d x=0,
\end{equation}
Moreover, (\ref{energy density relation4}) becomes
\begin{eqnarray}
&{\ }&\int(2x_tx_sK_s-x_sx_sK_t)d^d x=\int x_sx_s(x_t\alpha_{r,r}-b_t)\\ \nonumber
&=&\int(2x_tx_sK_s-x_sx_sK_t)d^d x+\int\{ x_t(2\beta_{rr}+2(d+2)\zeta-2\bar\zeta)+2x_s(\beta_{st}+\beta_{ts}-\bar\beta_{ts})\}d^dx,
\end{eqnarray}
Then, from this we get 
\begin{equation}
\int\{ x_t(2\beta_{rr}+2(d+2)\zeta-2\bar\zeta)+2x_s(\beta_{st}+\beta_{ts}
-\bar\beta_{ts})\}d^dx=0
\end{equation}
Finally we examine (\ref{energy density relation5}). (\ref{ansztz--apha-b}) satisfies this under a condition that
\begin{equation}
\int\{2x_tx_u(2\beta_{ut}-\bar\beta_{ut})+x_sx_s(2\beta_{tt}-\bar\beta_{tt}
+2(2+d)\zeta-(2+d)\bar\zeta+c_{uu}) \}=0.
\end{equation}

Minimal solutions of the coefficients in front of $\delta$-functions on the right hand side of (\ref{commutator-UUprime}) are given by
\begin{equation}
2\alpha_t=b_t=2K_t, {\ \ \rm and\ \ }\beta_{st}=\bar\beta_{st}=\zeta=\bar\zeta=c_{rs}...=0
\end{equation}
Therefore, the minimal solution of the commutation relation between the energy densities which satisfies conformal algebra becomes
\begin{equation}
\label{UU-commm}
[U,U^\prime]=K_{t,t}\delta+2K_t\delta_{,t}.
\end{equation}

\section{Conformal invariance and 4-$d$ vector theories}
\label{section2}
The main consequence of the last section is (\ref{UU-commm}). Once we quantize our vector field theory as (\ref{BB-commm}) and define the momentum and the energy densities as (\ref{em-density-intro}), then this satisfies\cite{Bunster:2012hm}
\begin{equation}
[U,U^\prime]=-\varepsilon\delta^{st}(K_s+K^\prime_s)\delta_{,t},
\end{equation}
where $\varepsilon=0 {\rm \ or }-1$. Conformal algebra is consistently constructed from the energy density only when the conformal dimension of energy density is  $\Delta_E=4$ in 4-dimensional spacetime. The simplest candidate for this is $U=h$, since $\mathcal B^a_s$ has conformal dimension $2$.

\section*{Acknowledgement}
We would like to thank Alfred D. Shapere for the useful discussion. J.H.O thanks his $\mathcal W.J$. 
This work is supported by the research fund of Hanyang University(HY-2013) only.



\end{document}